\documentclass[runningheads]{llncs}
\usepackage[T1]{fontenc}
\usepackage{todonotes}
\usepackage{graphicx}
\usepackage{tabularx}
\usepackage{booktabs}
\usepackage{url}
\newcommand{\code}[1]{\texttt{\scriptsize#1}}

\usepackage{listings}
\usepackage{xcolor}

\definecolor{keywordcolor}{rgb}{0.67,0.13,1.00}
\definecolor{commentcolor}{rgb}{0.00,0.50,0.00}
\definecolor{stringcolor}{rgb}{0.06,0.10,0.98}

\lstdefinelanguage{SPARQL}{
    keywords={PREFIX, SELECT, WHERE, FILTER, ORDER, BY, a},
    keywordstyle=\color{keywordcolor}\bfseries,
    sensitive=true,
    commentstyle=\color{commentcolor}\ttfamily,
    stringstyle=\color{stringcolor}\ttfamily,
    morecomment=[l][\color{commentcolor}]{\#},
    morestring=[b]",
}

\begin{document}
\title{The Climate Change Knowledge Graph:\\Supporting Climate Services}% in the Context of Climate Change}
%
%\titlerunning{Abbreviated paper title}
% If the paper title is too long for the running head, you can set
% an abbreviated paper title here
%
\author{Miguel Ceriani\inst{1}\orcidID{0000-0002-5074-2112} \and
Fiorela Ciroku\inst{1}\orcidID{0000-0002-3885-6280} \and
Alessandro Russo\inst{1}\orcidID{0009-0006-9437-0249} \and
Massimiliano Schembri\inst{1}\orcidID{0000-0001-9625-0052} \and
Fai Fung\inst{2}\orcidID{0000-0003-4367-151X} \and
Neha Mittal\inst{2} \and
Vito Trianni\inst{1}\orcidID{0000-0002-9114-8486} \and 
Andrea Nuzzolese\inst{1}\orcidID{0000-0003-2928-9496}
}
% \and
% Third Author\inst{3}\orcidID{2222--3333-4444-5555}}
%
\authorrunning{M. Ceriani et al.}
% First names are abbreviated in the running head.
% If there are more than two authors, 'et al.' is used.
%
% \institute{Princeton University, Princeton NJ 08544, USA \and
% Springer Heidelberg, Tiergartenstr. 17, 69121 Heidelberg, Germany
\institute{CNR - Institute of Cognitive Sciences and Technologies, Italy\\
\email{miguel.ceriani@cnr.it}, \email{fiorelaciroku@cnr.it}, \email{andrea.nuzzolese@cnr.it}, \email{alessandro.russo@cnr.it}, \email{massimiliano.schembri@cnr.it}, \email{vito.trianni@cnr.it}%\\
% \url{http://www.springer.com/gp/computer-science/lncs} \and
% ABC Institute, Rupert-Karls-University Heidelberg, Heidelberg, Germany\\
% \email{\{abc,lncs\}@uni-heidelberg.de}
\and
Met Office, UK\\
\email{fai.fung@metoffice.gov.uk}, \email{neha.mittal@metoffice.gov.uk}}

\maketitle              % typeset the header of the contribution
\begin{abstract}
Climate change impacts a broad spectrum of human resources and activities, necessitating the use of climate models to project long-term effects and inform mitigation and adaptation strategies. These models generate multiple datasets by running simulations across various scenarios and configurations, thereby covering a range of potential future outcomes. Currently, researchers rely on traditional search interfaces and APIs to retrieve such datasets, often piecing together information from metadata and community vocabularies. The Climate Change Knowledge Graph is designed to address these challenges by integrating diverse data sources related to climate simulations into a coherent and interoperable knowledge graph. This innovative resource allows for executing complex queries involving climate models, simulations, variables, spatio-temporal domains, and granularities. Developed with input from domain experts, the knowledge graph and its underlying ontology
are published with open access license
and provide a comprehensive framework that enhances the exploration of climate data, facilitating more informed decision-making in addressing climate change issues.
\keywords{climate change  \and knowledge graph \and climate service.}
\end{abstract}
\section{Introduction}

Climate change is a pressing global challenge, exerting profound effects on diverse aspects of human life and the planet~\cite{ipcc}. Its impacts permeate ecosystems, economies, and societies, necessitating robust strategies for mitigation and adaptation to safeguard their futures. Central to these strategies is the ability to accurately project the future climate through sophisticated models. These models serve as indispensable tools for policymakers and scientists, enabling them to anticipate changes, devise strategies, and implement responses to stabilize and potentially reverse climate impacts~\cite{pachauri}.

Various climate models have been developed globally, each employing different theoretical underpinnings and computational techniques.
%(Collins et al., 2006\cite{collins}).
They are driven by multiple standardized scenarios—such as those formulated by the Intergovernmental Panel on Climate Change (IPCC)—which reflect varying levels of greenhouse gas emissions and socio-economic developments.
%(Moss et al., 2010\cite{moss}).
Each climate model's output comprises multifaceted datasets representing a variety of variables over extensive spatial and temporal domains, necessitating data-intensive approaches to analyze the full spectrum of potential climate futures. %(Taylor et al., 2012\cite{taylor}).

Traditional data retrieval mechanisms, including conventional search interfaces and flat Application Programming Interfaces (APIs), have been the mainstay for accessing these outputs. However, these systems fall short in accommodating the complex, interrelated nature of climate data; they often require the integration of metadata and vocabularies to contextualize the information needed for comprehensive analysis. %This fragmented approach impedes the ability of researchers to efficiently derive meaningful insights and constrains interdisciplinary exploration (Hassani et al., 2020\cite{hassani}). 

To address these limitations, we introduce the Climate Change Knowledge Graph,
%a pioneering initiative in the field of climate science. This knowledge graph
which aggregates diverse climate-related data sources into an interconnected, interoperable structure, reflecting the complexity and depth of climate models, simulations, and their associated variables. By offering a unified framework capable of supporting intricate queries, the Climate Change Knowledge Graph facilitates %enhanced
data exploration and analysis, empowering experts to conduct more nuanced and sophisticated investigations. This project was developed through iterative dialogues with climate science domain experts, ensuring that the ontology and data representation effectively meet the community's needs.

The remainder of this paper delves into the construction and capabilities of the Climate Change Knowledge Graph. We analyse existing related work (Section~\ref{sec:related}), introduce the relevant methodologies employed in its development (Section~\ref{sec:meth}), define its scope and requirements (Section~\ref{sec:scope}), present the created resource (Section~\ref{sec:kg}), describe its functional evaluation (Section~\ref{sec:eval}), and offer some conclusions and future work ideas (Section~\ref{sec:conclusions}).
%By advancing the integration and accessibility of climate data, our work aspires to support more informed decision-making in climate change mitigation and adaptation strategies.

\section{Related Work}\label{sec:related}

%\todo[inline]{Add links to open source ontologies and other terms presented in the section}

The semantic representation and integration of climate data has attracted increasing attention in recent years, with several works proposing ontology-based frameworks to enable interoperability, reasoning, and decision support.
%The focus in the field has been on structuring heterogeneous climate data using formal ontologies to facilitate linked data publishing and semantic queries.
For example, Wu et al. in \cite{wu2022linkclimate} present an interactive knowledge graph platform, named \textit{LinkClimate}, that harmonises climate data using ontological modelling and linked data principles.
%In addition, the knowledge graph supports advanced data access and exploratory queries.
%This work can be seen as an extension of the authors' earlier work \cite{9553547}, in which they describe the \textit{Climate Analysis (CA)} ontology.
The purpose of the KG %ontology
is to represent datasets of climate observations, focusing on those from the National Oceanic and Atmospheric Administration.
% For the development of the CA ontology the authors reuse standard ontologies and vocabularies such as the \textit{Semantic Sensor Network} ontology, the \textit{WGS84} ontology for geographical locations, \textit{Climate and Forecast} vocabulary for geophysical properties, and the  \textit{QUDT} vocabulary for units of measurement. The \textit{CA} ontology is furthermore expanded to include a wider range of climatic and observational data, as presented in \cite{10282269}. This expansion allows a more comprehensive representation of the climate phenomena concepts. In both papers, the authors describe practical applications of the ontology in order to demonstrate its utility in real-world scenarios. 

In more specialized contexts, several domain ontologies have been proposed to model climate-related knowledge. %, with a focus on knowledge representation and data integration.
%Another example of an ontology is presented in
Naidoo et al.~\cite{9519380} %.The authors
describe the \textit{Climate Smart Agriculture} ontology, which addresses the need for representing knowledge related to agricultural practices in the face of climate change. Furthermore, Surya et al. \cite{10.1007/978-3-030-71187-0_104} propose a climate change ontology which takes into consideration the social and geological aspects, this way emphasising the interchange between natural phenomena and human-related factors. Ma et al. \cite{MA2014191} offer an alternative viewpoint on the subject by developing a provenance ontology for climate assessment. The purpose of the ontology is to ensure the transparency and traceability of the U.S. Global Change Research Program's National Climate Assessment reports by allowing the users to track the information flow between data sources, agents, and conclusions made. In other cases such as in Lippolis et al. \cite{lippolis2025water}, ontology networks have been developed for integrating meteorological, health, and environmental data into a single source. The \textit{Health Water Open Knowledge Graph} can be used for a wide range of purposes in the water and health domains starting from knowledge discovery to policy-making. Lastly, in the context of policy making and management, Kontopoulos et al. \cite{kontopoulos2018ontology} formalize an ontology which represents crisis management procedures for extreme weather events.
%The ontology enhances analytical methodologies by highlighting situational semantics and procedural interaction.
Together, these contributions demonstrate the growing importance of ontology engineering and knowledge graph development in climate-related fields.

Our work is the first to directly address the representation of model-based climate projections, their properties, and relationships. The analysis of climate projections is the main tool climate scientists and practitioners have in order to gain insights on the climate of the next decades and try to prepare accordingly.

% Although the domain of the fore-mentioned ontologies is not climate science directly, it is a step forward on considering climate information as crucial in agriculture, health, geology crises management, water management, and other sectors. Our work builds on this foundation by providing ......, further supporting .....

\section{Methodology}\label{sec:meth}

This section describes the methodologies adopted to create the knowledge graph,
specifically the methodology for ontology design and the general workflow to incorporate existing knowledge.

\subsection{Ontology Design}\label{sec:meth-xd}

Extreme Design (XD) is an ontology design methodology that is characterized by the reuse of Ontology Design Patterns (ODPs), modular design and a test-driven approach as guiding principles \cite{presutti2009extreme}. The reuse of Ontology Design Patterns, which serves both as a principle and as a development practice, offers an approach to address recurring modeling issues. In essence, ODPs are ontologies that act as a bridge between use cases (types of problems) and design solutions. They are used as modeling components: an ontology should ideally be the result of a composition of ODPs with appropriate dependencies between them, as well as the necessary design extensions based on specific needs \cite{gangemi2009ontology}. ODPs can be found in catalogs, such as the Ontology Design Patterns Portal\footnote{\url{http://ontologydesignpatterns.org/wiki/Main_Page}}, the Ontology Design Patterns Workshop Series\footnote{\url{http://ontologydesignpatterns.org/wiki/WOP:Main}}, and the University of Manchester catalog\footnote{\url{http://www.gong.manchester.ac.uk/odp/html/}}. Whereas, the principle of modular design consists in separating the modeling of the requirements into independent, interchangeable modules. Each of the modules contains everything needed to satisfy even just one component of the desired requirement. Additionally, XD focuses on ontological unit testing, making it analogous to software testing, while other ontology modeling methodologies involve testing of a more semantic nature.

\subsection{Incorporating Existing Knowledge}\label{sec:meth-map}

\begin{figure}[htbp]
    \centering
 %   \begin{minipage}{0.9\textwidth}
% PREFIX rdfs: <http://www.w3.org/2000/01/rdf-schema#>
% PREFIX top: <https://w3id.org/hacid/onto/top-level/> 
% PREFIX ccso: <https://w3id.org/hacid/onto/ccso/>
% PREFIX data: <https://w3id.org/hacid/onto/data/>
% PREFIX rcp: <https://w3id.org/hacid/data/cs/greenhousegasconcentrationpathway/>
% PREFIX mip: <https://w3id.org/hacid/data/cs/variable/mip/>
% PREFIX dim: <https://w3id.org/hacid/data/cs/dimension/>

%store them as local files;
%    ####### 1. DATA CLEANSING #######
%    ####### 2. MAPPING #######
%    reformat (if needed) to help mapping (e.g., JSON to CSV);
%    ####### 1. DATA CLEANSING #######
%    ####### 2. MAPPING #######

    \lstset{language=SPARQL, frame=single, basicstyle=\ttfamily\small}
    \caption{Process adopted to integrate existing data-sources into the KG}
    \label{lst:handle}
    \begin{lstlisting}
prepare an empty dataset on a local triple store;

for each datasource:
    gather structured or semi-structured data;
    perform data cleansing (if needed);
    map to RDF;
    upload RDF to local triple store;

for each consolidation task:
    run update on the local triple store;

export local dataset as RDF;
publish novel/updated RDF dataset on public triple store;
    \end{lstlisting}
 %   \end{minipage}
\end{figure}

In order to map existing structured data to RDF (and precisely to the defined ontology), we adopted a three-step data handling process: first data cleansing, then mapping, and finally consolidation. Such a multi-step paradigm enables us to solve issues that are cumbersome or unfeasible to be dealt with by a mapping language alone. 
Figure~\ref{lst:handle} shows a high-level view of the adopted process, described with pseudocode.

For every data source and every step of the process we automatised the procedure as much as possible and adopted existing standards.
%, specifically declarative mapping/transformation languages.
The main technology adopted for mapping to RDF is the  RDF Mapping Language (RML), which allows for multiple input formats (CSV, JSON, XML).
%RML itself is expressed in RDF, therefore potentially allowing inclusion in knowledge graphs for interoperability and full integration with the mapped data. % (e.g., to operationally represent provenance).
% Furthermore, RML offers an extension point in the possibility to include user defined functions, defined in some host language% (in our case, in Python);
% %while Python
% \footnote{Using host language functions reduces the reusability of the RML mapping,
% %in other languages,
% but they are often necessary to 
% %achieve some of the required
% non-trivial
% transformations.}.
For some data sources, data cleansing is needed before the actual mapping.
Some cleansing can be
%is
dealt by hand, %(if it is a very simple change),
while for
other cases scripts need to be developed (declarative language are preferred for the purpose, e.g., jq language to transform JSON).
% some sources in the JSON format jq is used. The jq language is a declarative language broadly adopted to perform JSON to JSON transformations.
% Furthermore, as RDF can be represented in JSON as JSON-LD, jq can be used also to perform the mapping from JSON to RDF (by generating JSON-LD).
% In some cases this method gives a more concise representation of the mapping and it is hence adopted for that purpose.

The consolidation tasks, finally, are additions to the knowledge graph that can be performed only after the mapping because they require some form of aggregation of the mapped data. They are represented as SPARQL updates. %of the form INSERT {...} WHERE {...}.

\section{Scope and Requirements}\label{sec:scope}

%\todo[inline]{I wrote down schematically the current structure of the section for our convenience, must be replaced later by an intro of the section.}

This Section introduces scope (sections~\ref{sec:ctxt-proj}, \ref{sec:ctxt-service}, and \ref{sec:ctxt-hacid}),
competency questions (Section~\ref{sec:reqs}), and data sources (Section~\ref{sec:sources}) for the development of the knowledge graph.

% \begin{itemize}
%     \item \ref{sec:meth}, description of the methodology(ies) in general;
%     \item \ref{sec:ctxt-proj}, \ref{sec:ctxt-service}, and \ref{sec:ctxt-hacid}, description of the context and main concepts;
%     \item \ref{sec:reqs}, description of the more detailed requirements (e.g., CQs) possibly referring to external sources and choosing few examples;
%     \item \ref{sec:sources}, description of structured data sources, alongside relevant infrastructures, formats, and conventions.
% \end{itemize}

%\todo[inline]{I would add here a very brief description of XD and other relevant methodologies applied to the specific case, both for ontology design and (if applicable) for kg creation.}

\subsection{Climate projections}\label{sec:ctxt-proj}

In order to study climate change, \emph{climate models} are used to project how the weather will behave in the foreseeable future, based on estimates of the effects human activities (mainly in terms greenhouse gas emissions).
Climate models simulate physical processes with some degree of approximation. While we (as humans) are mostly interested in the effects at the Earth's surface, climate models include both multiple atmospheric layers and subsurface (ocean or underground) layers that effects the whole climate system.
The output of such simulations that explore the future are called \emph{climate projections}.
They typically simulate the climate for a period of time in the future (e.g., the next 100 years) based on the historical evolution and current state, but they can also be run for a period entirely in the past (e.g., last 50 years) in order to evaluate the climate model and other assumptions.
Climate projections consist in practice in datasets containing values of a set of dependent variables (e.g., atmospheric temperature at the surface) for all the combinations of the independent variables (typically space and time, within the ranges considered in the simulation and discretized according to the required resolution).

Estimates on greenhouse gas emissions are, in turn, generated by socio-economic models guided by broad assumptions called \emph{emission scenarios}.
Widely used sets of emissions scenarios include the \emph{representative concentration pathways (RCP)} --- originally being \emph{RCP2.6}, \emph{RCP4.5}, \emph{RCP6}, and \emph{RCP8.5} to which more were added  later --- that are labelled after the expected changes in radiative forcing values from the year 1750 to the year 2100 (2.6 W/m$^2$, 4.5 W/m$^2$, and so on)
and the \emph{shared socioeconomic pathways (SSPs)} --- numbered from 1 to 5 --- that represent alternative storylines regarding socioeconomic development (according to global economic equality and sustainable growth)%
\footnote{Emission scenarios from complementary classifications can potentially be combined and as a fact RCPs and SSPs are also used in combination, forming a matrix of cases.}.

With regard to all these assumptions, there is a great amount of uncertainty and they have a significant impact on the outcome of the simulations.
In addition, the approximations adopted in climate models have a significant impact on how realistic the macroscopic climate processes can be represented.
For these reasons, there is more interest in comparing and contrasting multiple projections (obtained with different models and assumptions) rather than establishing a single ``correct'' projection.
\emph{Model intercomparison projects (MIPs)} provide frameworks to compare different models and evaluate the diversity of outcomes. They are organized in \emph{experiments}: each experiment is a set of simulation constraints as the considered emission scenario, the spatio-temporal range and resolution, the climate variables required in the output.
Multiple organisations contribute to MIP experiments by running simulations using their climate models with the constraints imposed by the corresponding experiment.

A well known MIP is the \emph{Coupled Model Intercomparison Project (CMIP)}, 
a project of the World Climate Research Programme (WCRP). It includes both experiments concerning time periods in the path and time periods in the future, considering different emission scenario. The space coverage is in all cases the whole Earth, to allow for comparison at a global level.
It is on-going process since 1995, organised in consecutively numbered \emph{phases}, the last completed one being CMIP6.

Regarding projections, the space-time resolution required may vary a lot according to the purpose and context of a study. Global climate models are typically run at a spatial resolution of 60-100 km and a temporal resolution ranging from 30 minutes to a few hours. To get data at lower resolution, especially on the temporal axis, the variable values are aggregated using some function (typically the mean, but also min and max). To obtain data at higher resolution, especially on the spatial axis, additional methods have to be adopted and the process is called \emph{downscaling}\footnote{It should be noted that in climatology downscaling is a process by which the resolution of a dataset is increased, whilst in other domains, like image processing, this would be called upscaling.}.
A special case is when, in order to increase the resolution, a higher resolution, more detailed climate model is run using the output of the original one as a constraint. Such fine-grained climate models usually focus on a limited region of the world and can be qualitatively different from global climate models. The process in this case is called \emph{dynamical downscaling} and the fine-grained models are called regional climate models or local climate models depending on the granularity.

A special case of aggregation of variable values is the computation of \emph{indices}, also known as \emph{metrics}, that are used to quantify the climate change is respect to specific hazards or conditions (e.g., the number days in a year in which the temperature is greater than or equal to 35 °C).

\subsection{Climate service}\label{sec:ctxt-service}

%\todo[inline]{I would add here a brief description of what is a climate service, a climate service provider, and an overview of the process.}

Climate services refer to the provision of climate information and related services that assist in decision-making across various sectors. These services aim to translate complex climate data into useful, actionable information that can inform policy, planning, and operational decisions to manage risks and exploit opportunities associated with climate variability and change~\cite{hewitt2012developing,vaughan2014climate,brasseur2016remaking}.
%The goal is to support climate adaptation and resilience by making climate science more accessible and applicable to end-users such as governments, businesses, and communities.

Through a collaborative process involving domain experts, we elicited the typical process that is followed in order to deal with a climate case.
% The process consists of five main phases:
% \begin{enumerate}
%     \item \emph{gather organisation background information}, i.e., get information on the requesting organisation regarding the sector they operate on, their needs, and goal;
%     \item \emph{gather climate case information}, i.e. get information on exposure and vulnerability, relevant hazards, risk tolerance, climate case description, required spatio-temporal coverage; 
%     \item \emph{prepare climate information}, i.e., select available climate projections based on spatio-temporal constraints for coverage and resolution, required variables, appropriated scenarios; possibly extend that information as needed (e.g., downscaling projections if more detail is necessary); compute relevant metrics from the projections;
%     \item \emph{analyse climate information}, by quantifying climate change and its uncertainty;
%     \item \emph{communicate climate information}, by offering relevant and actionable insights to the requesting organisation in the requested format, i.e. report, presentation, chart.
% \end{enumerate}
From the point of view of accessing to existing climate knowledge,
one of the most relevant tasks of the process
%, from the poin
is the selection of climate projections that are meaningful to addressing the specific case.
% Phase three is the one in which access to existing climate knowledge is most relevant. This phase has an initial sub-phase in which available climate projections are selected.
This is done by filtering through some criteria including the following:
\begin{itemize}
    \item by available (dependent) variables;
    \item by baseline period;
    \item by future period;
    \item by considered emission scenario.
\end{itemize}

% \todo[inline]{Move the refs below to bib}

% One key article by Hewitt, Mason, and Walland (2012), published in the journal Nature Climate Change, discusses the principles and development of climate services. They emphasize the importance of integrating scientific climate data with the decision-making processes of users to improve societal outcomes.

% Brasseur and Gallardo (2016) explore the evolution of climate services in their paper published in Science. They highlight the need for robust frameworks that facilitate the accessibility and usability of climate data.

% Vaughan and Dessai (2014) provide a conceptual analysis in Climate Risk Management, focusing on the decision-making processes inherent in climate services. They highlight the interdisciplinary nature of climate services and their dependency on active user participation.

\subsection{The HACID project}\label{sec:ctxt-hacid}

%\todo[inline]{I would add here a brief description of the HACID project, the role of collective intelligence, how the climate service process is augmented through collaboration (climate questions and contributions), and the role of the kg in this collaboration (both supporting the exploration/query of resources and their unambiguous identification and referral).}

This knowledge graph has been developed in the context of HACID\footnote{\url{https://www.hacid-project.eu/}}, a European project that aims to develop hybrid collective intelligence methods for decision support in multiple contexts.
The use case that is relevant here focuses on improving existing climate services by favouring collaboration between experts.
In the context of HACID, the knowledge graph is both a source of knowledge and a common vocabulary (here intended in a broad sense) to support the collective solution of climate cases.

\subsection{Competency Questions}\label{sec:reqs}

%\todo[inline]{Here more precise requirements (as CQs or similar) should be described, possibly referring to external sources and choosing few examples (maybe one example fully developed and other sketched).}

Based on the types of information described in Section~\ref{sec:ctxt-proj} and the usage context (for climate services) described in Section~\ref{sec:ctxt-proj}, competency questions (CQs) have been elicited throughout the whole design process.
The current version of the set of CQs is available publicly%
\footnote{\url{https://github.com/hacid-project/climate-services-kg/blob/main/doc/competency-questions/list.md}}.
This resource is continually updated and extended as new requirements arise. The corresponding SPARQL queries are shown alongside the CQs, both for evaluation purposes and to provide examples of use. 
The following are two example CQs.

\begin{sloppypar}
\begin{itemize}
    \item \emph{\textbf{Filter simulations} Filter available simulations by chosen emission scenario (e.g., RCP-4.5), available variables (e.g., tas), and required spatial resolution (e.g., less than 0.2 degrees).}
    \item \emph{\textbf{Filter output datasets} 
Filter datasets from a simulation (e.g., cmip5.HadCM3.rcp45.r10i1p1), based on available variables (e.g.,  tas), and required exact temporal resolution (e.g., one month)}.
\end{itemize}
\end{sloppypar}

% \subsection{Data Infrastructures}\label{sec:infra}

% In order to integrate existing structured knowledge, data infrastructure and formats of the domain have been explored.
% The following 

\subsection{Data Sources}\label{sec:sources}

% \todo[inline]{The part below was copied from a non-final version of D2.2, because I thought there was some interesting things to recover, but we may as well delete all this part and leave the newer one}

%A collection of relevant structured data sources has been identified through continued interaction with domain experts.
A view of the data landscape of the domain has been obtained through continued interaction with experts.
Both data sources, data infrastructure, conventions, and formats have been explored, including the ones described below, deemed crucial for the KG development.
The core of the information of interest is about climate simulations, their setup (e.g., the adopted climate model or the considered emission scenario) and their outputs (including how they are organized in datasets). Table~\ref{tab:climate-components} provides an overview of the main external components integrated into the knowledge graph, highlighting their roles and relevance in representing climate data and metadata.

\begin{table}[ht]
\centering
\caption{Summary of data sources used for constructing the knowledge graph.}
\label{tab:climate-components}
\resizebox{1\textwidth}{!}{
\begin{tabular}{p{5cm}p{10cm}}
\toprule
\textbf{Data source} & \textbf{Description} \\
\midrule
\textbf{NetCDF (Network Common Data Form)} & A self-describing, machine-independent format for array-oriented scientific data, widely used in climate datasets. In this project, only the metadata is accessed and represented in the knowledge graph to describe simulations and their properties, not the full data content. \\
\textbf{ESGF (Earth System Grid Federation)} & A distributed infrastructure for managing and accessing climate model and observational data. ESGF is used as a metadata source and access point for climate datasets relevant to the project. \\
\textbf{CF Metadata Conventions} & A set of standards for describing Earth science data in NetCDF files, defining variables and their spatial/temporal properties. These conventions support consistent metadata interpretation in the knowledge graph. \\
\textbf{CMIP CVs and CMOR Tables} & Controlled vocabularies and tables that describe key metadata elements (experiments, variables, models, etc.) used in CMIP datasets. The CMIP6 CVs are represented in the knowledge graph to support semantic indexing and integration. \\
\textbf{CMIP Datasets} & Outputs of climate model simulations structured by experiment, model, and configuration. A selected subset of CMIP5 datasets (historical and RCP scenarios) is mapped to the knowledge graph to represent both the datasets and their underlying simulations. \\
\textbf{CORDEX} & A framework for regional climate downscaling using domain-specific models and grids. CORDEX data and domain specifications are semantically represented, complementing the global CMIP5 datasets in the knowledge graph. \\
\textbf{Climdex} & A global collaborative project providing indices that quantify climate-related hazards using core climate variables. These indices are included in the knowledge graph as derived indicators relevant to impact assessment. \\
\bottomrule
\end{tabular}
}
\end{table}

\section{Resource: the Knowledge Graph}\label{sec:kg}

The knowledge graph is publicly accessible through both a SPARQL endpoint%
\footnote{\url{https://w3id.org/hacid/cs/sparql}}
and a linked data based interface (offering either human-readable content or RDF serializations, thanks to content negotiation) that can be reached by dereferencing any URI of the KG%
\footnote{As an example, \url{https://w3id.org/hacid/data/cs/variable/mip/tas}}.
Furthermore snapshots of the content of the KG are published on an open access repository for persistence%
\footnote{\url{http://dx.doi.org/10.6084/m9.figshare.29066711}}.
It is currently composed of around 14 million triples organized in 13 named graphs (used for provenance, when needed%
\footnote{The default graph is setup to be the union of all the named graphs.}).
The data is provided under the Creative Commons Attribution 4.0 International license%
\footnote{\url{https://creativecommons.org/licenses/by/4.0/}}, hence by allowing broad reusability.
Finally, the used ontologies are also published online, both as OWL and as web pages with human-readable description of the terms.
The rest of the section describes the ontology network adopted and the data contained in the KG.

\subsection{Ontology Network}

The ontology network is composed of three main modules:
\begin{itemize}
    \item \emph{top-level}\footnote{\url{https://w3id.org/hacid/onto/top-level} associated with prefix \code{top:}}, providing general common terms for the whole HACID project (see Section~\ref{sec:ctxt-hacid});
    \item \emph{Core Climate Services Ontology (CCSO)}\footnote{\url{https://w3id.org/hacid/onto/ccso} associated with prefix \code{ccso:}}, specific to the domain; and
    \item \emph{data}%
    \footnote{\url{https://w3id.org/hacid/onto/data} associated with prefix \code{data:}}, an ontology to represent variables, dimensional spaces, and datasets.
\end{itemize}

In the following, the ontology network is described by functional components.
When not explicitly marked otherwise, the described components are part of CCSO.

\paragraph{\bf Common Terms (top-level).}

The top-level ontology is designed to be at the level of an upper-ontology, to provide common general terms to be used in the HACID project (that is, not only for the knowledge graph described here).
A part from a few exceptions the top-level ontology is a subset of DOLCE ultra light and for that reason it is not described here, leaving space to the other, more specialized, modules.

\paragraph{\bf Climate Models.} A central aim of climate science is trying to project future climate conditions, based on past
and present data and adopting some model for the considered system. The models (class
\code{\code{ccso:Model}}) considered here are algorithms that represent the behaviour and evolution of a
system in time. The models considered are often climate models (\code{ccso:ClimateModel}), i.e.,
representing climate dynamics in physical terms, but socioeconomic models like IAMs
(Integrated Assessment Models, class \code{ccso:IntegratedAssessmentModel}) are also of
interest when dealing with future collective human behaviour\footnote{Integrated assessment models are represented in the ontology but not currently incorporated in the KG}.
Different climate models are used for different spatial coverages/granularities: a
\code{ccso:GlobalClimateModel} is one that encompasses the whole earth surface, usually with
coarse granularity (e.g., 60 Km); a \code{ccso:RegionalClimateModel} is one that encompasses
only a part of the world (e.g., Europe) but is more fine grained (e.g., 12 Km); a
\code{ccso:LocalClimateModel} is one that encompasses a yet smaller area (e.g., the United
Kingdom), with still more detail (e.g., 2 Km). A special case of local climate models are
convection permitting models (class \code{ccso:ConvectionPermittingModel}), i.e. models with
enough detail to explicitly represent convection.

\paragraph{\bf Climate Simulations.}

\begin{sloppypar}
The execution of a model is a simulation (\code{ccso:ModelBasedSimulation}), as it simulates the behaviour of some (real)
system.
Climate model simulations generate as output projections, which are datasets.
% which are, as the emission scenarios, spatiotemporal datasets (further described below).
A specific case of a \code{ccso:ClimateModelSimulation}
%(not being a \code{ccso:GCMSimulation})
is a \code{ccso:DynamicalDownscaling}, i.e. when a simulation is constrained by the output of an
upscale simulation (e.g., a projection obtained by a global climate model simulation is used
as input to a regional climate model simulation).
% \code{ccso:ProjectionDownscaling} is the
% generalisation of \code{ccso:DynamicalDownscaling} to processes that are not simulations, for
% example the downscaling of a projection by statistical means
% (\code{ccso:StatisticalDownscaling}).
An important aspect for a climate simulation is the chosen emission scenario (\code{ccso:EmissionScenario}) which can be defined in multiple ways. Two specific classes address two of the most common classifications: shared socio-economic pathways (SSPs) and representative concentration (of greenhouse gas) pathways (RCPs).
The property \code{ccso:refersToScenario} allows the direct association of a climate simulation with the considered emission scenario.
\end{sloppypar}

\begin{figure}[ht]
    \centering
    \includegraphics[width=\textwidth]{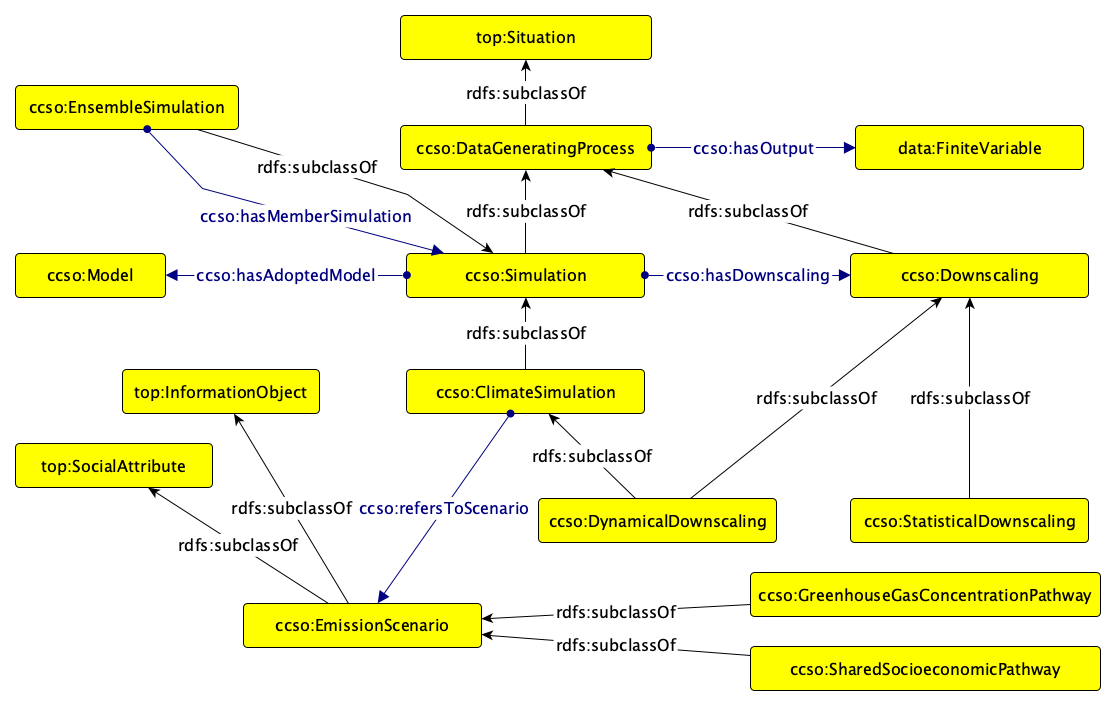}
    \caption{Graffoo diagram depicting the part of ontology related to climate simulations.}
    \label{fig:onto-simulations}
\end{figure}

\paragraph{\bf Variables, Dimensional Spaces, and Datasets (data).}
Based on the requirements, we elicited the need for a data-centric representation, based on the properties of datasets in terms of variables (which may be dependent from other variables or not, computationally derived from other variables or not) and the dimensional spaces they take values in (which can be discrete or continuous, made of points or extended regions, limited or not, derived-from/part-of other dimensional spaces).
%For that purpose a dedicated submodule of CCSO has been developed, associated with the namespace \code{https://w3id.org/hacid/onto/data}, here shortened with the prefix \code{data:}.

The aim is to represent the concept of variables in a flexible way, while preserving well-defined semantics. Here, as usual in mathematics and other fields, a variable is an abstraction that has meaning in the context of some relation between different variables and can be associated to different values in a set that is usually well defined (e.g., $x\in\mathbf{R}$). In the ontology module presented here, a \code{data:Variable} is akin to that concept and the set of possible values is represented by a \code{data:DimensionalSpace}; the two are linked via the \code{data:hasValuesOn} property.
A variable that is a \code{data:DependentVariable} can be implicitly associated with a function having as range the associated dimensional space and as domain the cartesian set of the dimensional spaces of the variables from which this variable depends. A \code{data:IndependentVariable}, conversely, is a variable that does not depend on other variables.
A variable can be \emph{aggregating}, i.e. include aggregation mechanisms whose execution is deferred to derived variables. This allows to represent variables defined using one or more aggregation functions, %for combining data along one or more independent variables,
without specifying the actual granularity.
%A general purpose example of usage is representing as variables the SQL expressions including aggregation functions (e.g. AVG(...)) which are then derived to actual serialisable variables by the usage of specific GROUP BY clauses.
% In the context of climate services, the CMOR Tables variables (e.g., tasmax) fall into this category. From the ontology point of view, a (dependent) variable is aggregating if it offers one or more aggregations: i.e. it is linked with one or more instances of  \code{data:Aggregation} via the property \code{data:definesAggregation}. Each aggregation is associated with one variable (which must be a variable the dependent variable depends on) and optionally one or more suggested options for the adopted aggregation grid (i.e. specific granularity).

A \code{data:FiniteVariable} is a variable for which the set of all the possible values is finite. For an independent variable it means being defined on a finite dimensional space (instance of \code{data:FiniteDimensionalSpace}); for a dependent variable it means that all the variables it depends on are finite variables. The concept of \code{data:FiniteVariable} is important because it corresponds to variables that can be fully represented extensively with a finite amount of resources. It corresponds functionally to the concept of a dataset. The property \code{data:hasDataSerialisation} links such a dataset with a data source (\code{data:DataSource}) holding one or more concrete serialisations of it. A \code{data:DataSource} (e.g., an URL resolving to a dump of a dataset) is associated to one or more supported data formats (\code{data:DataFormat}) via the \code{data:hasAvailableDataFormat} property. 

\begin{figure}[ht]
    \centering
    \includegraphics[width=\textwidth]{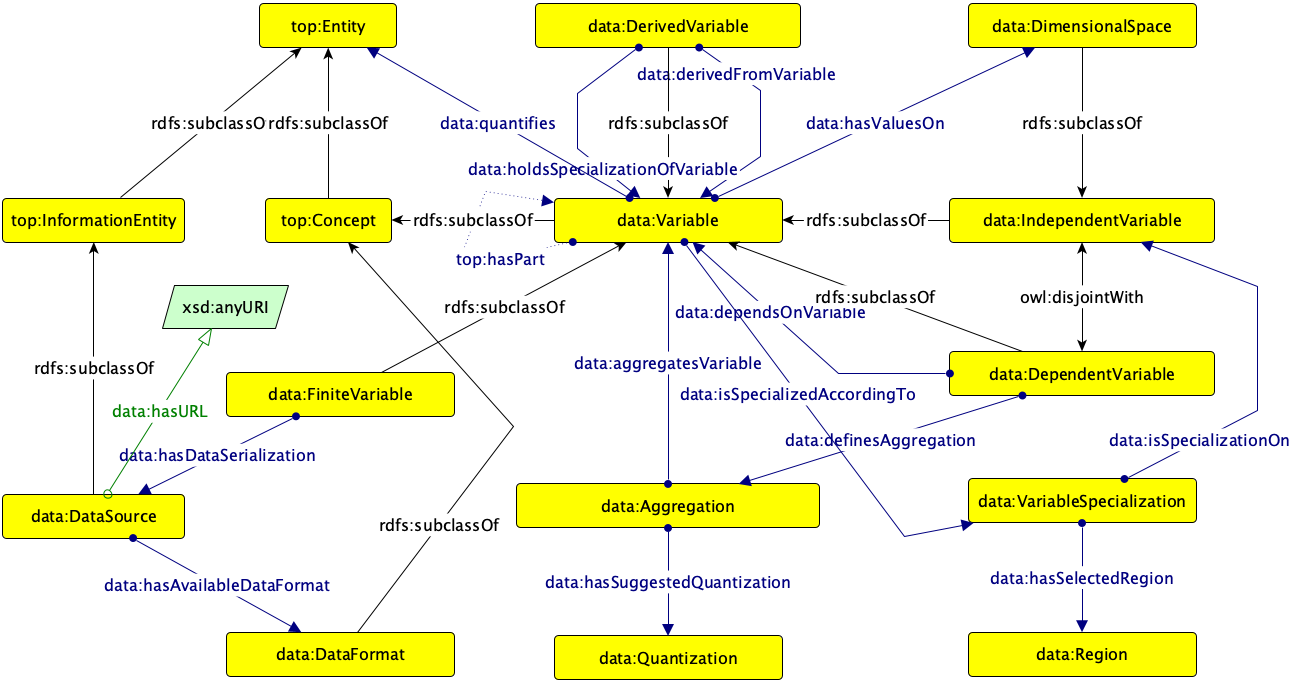}
    \caption{Graffoo diagram depicting the part of ontology related to variables.}
    \label{fig:onto-variables}
\end{figure}

\begin{sloppypar}
A \code{data:DimensionalSpace} is composed of a set (not necessarily finite) of regions (\code{data:Region}). It can be either continuous (\code{data:Continuum}) or discrete (\code{data:DiscreteDimensionalSpace}). A discrete dimensional space can be a grid (\code{data:Grid}), i.e. composed of a set of regions that form a partition of all the space.
In order to avoid unnecessary complexity in concrete usage, a \code{data:DimensionalSpace} is also a \code{data:IndependentVariable}. Conceptually, it associates a dimensional space with a “default variable” with values on that dimensional space (on itself). This allows using directly a dimensional space as a variable when there are no possible ambiguities: e.g., a variable representing time in a context where no other variables represent time, can be directly represented by the corresponding temporal dimensional space.
\end{sloppypar}

\subsection{Data Content}

The content of the KG is built upon a diverse array of data sources (cf. Table~\ref{tab:climate-components}), each contributing specialized information. The resources have been already presented in Section~\ref{sec:sources}, in the following we detail what and how data are incorporated in the KG.

\paragraph{\bf CF Metadata Conventions.}
%are relevant on their own and as a basis of the conventions used in the context of CMIP.
All the 4,872 standard variables defined in CF are incorporated in the KG, alongside the description and their unit of measure.

\paragraph{\bf CMOR Tables} describes two types of variables: 1,273 Model Intercomparison Project (MIP) variables, derived from CF variables by specializing them and specifying aggregation methods; and 2,068 CMOR variables further specialized by choosing a time granularity.
As an example, the CF variable \code{air\_temperature} is specialized by the MIP variable \code{tasmax} in two ways: it is taken near-surface and when aggregated in time the maximum is used (rather than the mean, minimum, or other aggregation functions).
The MIP variable \code{tasmax} is further specialized by the CMOR variable \code{tasmax.mon} that specify a temporal granularity (monthly).
All the MIP variables are incorporated in the KG and the property \code{data:isSpecializationOfVariable} allows to maintain the association with the corresponding CF variables.
The CMOR are not mapped explicitly since they are redundant when used to describe datasets, whose time granularity can be described independently and once for all the MIP variables used.
%This dataset is mapped through RML from the source, which is available in JSON format via Git.

%\paragraph{\bf CMIP Controlled Vocabularies (CVs).}
\paragraph{\bf CMIP CVs.}
%define unique identifiers and labels for institutions, models, model intercomparison projects (MIPs), and MIP experiments.
Information
%about
on
institutions and models is incorporated in the KG.
%Description of MIPs and experiments is not currently part of our scope (as those describe the comparison framework, rather than the properties of simulations and models).

\paragraph{\bf CMIP5 Datasets} are generated under the Fifth Coupled Model Intercomparison Project (CMIP5).
% , represent a %comprehensive 
% collection of climate model outputs contributed by multiple institutions around the globe.
The meta-data associated with the datasets allow to identify the used climate model, the constraints (e.g., the reference emission scenario), the variables, the spatio-temporal coverage, and the spatio-temporal granularity.
While this set of meta-data is in itself flat, the mapping to the KG allows to give it depth:
rather than representing just the specific datasets and their properties, simulations and their outputs are represented.
A simulation, % is
identified by a specific model, a specific set of constraints (e.g., a specific reference emission scenario), and a specific configuration setup,
%. In the KG it
becomes an instance of \code{ccso:GlobalClimateSimulation}.
%A simulation
It can have multiple outputs, each one identified as the simulation plus a specific label (e.g., \code{output1}, \code{output2}, ...),
associated via the
%The
property \code{ccso:hasOutput}.
%is used to associate the simulation with its outputs.
An output correspond in turn %general
to multiple datasets, typically offering different variables or different time granularities.
%Conceptually,
As conceptually
an output is already a dataset,
%For that reason
both outputs and single datasets are represented in the KG as instances of \code{data:Dataset},
hierarchically organized by
% . The outputs are connected to the corresponding single datasets with the property
\code{top:hasPart}.
The data currently incorporated in the KG correspond to datasets from a specific subset of experiments, namely the main future scenarios with representative concentration pathways (RCPs) and the main historical experiment.

% The datasets are then parts of the outputs, connected with the \code{top:hasPart} property.
% aimed at understanding past, present, and future climate changes.
% These datasets encompass simulations from multiple models, covering scenarios such as historical climate trends, future projections based on greenhouse gas concentration pathways, and idealized experiments.

% With over 6,000 datasets and 150+ simulations, CMIP5 provides standardized, high-resolution outputs across various spatial and temporal scales. Accessible via the Earth System Grid Federation (ESGF), the datasets conform to strict metadata and data format standards, enabling seamless integration and comparison. CMIP5 has been instrumental in advancing global climate assessments, including those by the Intergovernmental Panel on Climate Change (IPCC), supporting research on climate variability, impacts, and policy formulation.
% Accessed via the ESGF API, this dataset includes 152 simulations and a vast repository of 6,365 individual datasets, all delivered in JSON format.

\paragraph{\bf CORDEX Domains} includes 14 %different
geographical regions associated to 65 geodetic grids (covering variations in %the
region specification%
\footnote{Each region can be represented in standard Earth coordinates or in rotated polar coordinates (which ensure a more uniform representation in the latitude-longitude-based-grid).}
and multiple resolutions).
 % Furthermore, grids with different resolutions are defined.
\paragraph{\bf CORDEX Datasets} are treated similarly to the CMIP5 Datasets case. In addition to the method described for CMIP5, in this case the meta-data allows also to link in the KG each dynamical downscaling to the original simulation, via the property \code{ccso:isDownscalingOf}.
\paragraph{\bf Climdex.}
%includes 63 indices grouped into 9 categories. They 
The indices are described analogously to other variables, using the property \code{data:derivedFromVariable} to associate each index with the (MIP) variable it is computed from.

\subsection{Implementation}

The general implementation characteristics are given below for each of the main components.
The reader is referred to the %two
corresponding
%following
%mentioned
public git repositories for further detail.
% \begin{itemize}
%     \item \code{knowledge-graph}%
%     \footnote{\url{https://github.com/hacid-project/knowledge-graph}},
%     for design and development of the ontology network%
%     \footnote{Alongside code related to the application semantic technologies to other aspects of the HACID project.};
%     \item \code{climate-services-kg}%
%     \footnote{\url{https://github.com/hacid-project/climate-services-kg}},
%     for the development of the knowledge graph (specifically code for the mapping of existing data sources).
%     \item \code{climate-services-kg-docker}%
%     \footnote{\url{https://github.com/hacid-project/climate-services-kg-docker}},
%     for the docker-based deployment of the knowledge graph.
% \end{itemize}

\paragraph{\bf Ontology.} Ontology modules are implemented in OWL with a combination of Protégé and SPARQL queries/updates on the RDF representation.
The usage of reasoners in Protégé and check up queries aids in keeping the ontology consistent and maintain other desired quality features, as the appropriate annotation of terms.
For further detail check the online repository%
\footnote{\url{https://github.com/hacid-project/knowledge-graph}}.

\paragraph{\bf Data Mapping.} All the code used to map the data to the climate services knowledge graph is publicly available as part of the GitHub repository with code related to the HACID ontologies and knowledge graphs.
The code (a part from the declarative bits in RML, jq, and SPARQL Update) is mainly Python, using pyRML for execution of RML mappings and other libraries for specific data conversions (e.g., managing dates and times).
In addition there are some shell scripts that interact with the triple store or execute jq mappings. The triple store adopted for the mapping phase is Apache Jena Fuseki.
For further detail check the online repository%
\footnote{\url{https://github.com/hacid-project/climate-services-kg}}.

\paragraph{\bf Deployment.} The public-facing server is setup as a set of dockerized components:
%\footnote{See docker compose configuration at \code{https://github.com/hacid-project/climate-services-kg-docker/blob/main/docker-compose.yaml}}:
\begin{itemize}
    \item a GraphDB\footnote{\url{https://www.ontotext.com/products/graphdb/}} instance holding the triple store and offering SPARQL access;
    \item a LodView\footnote{\url{https://github.com/LodLive/LodView}} instance providing linked data (LD) access on top of the SPARQL endpoint;
    \item a nginx\footnote{\url{https://nginx.org/}} instance used to expose both the SPARQL endpoint and LD access on appropriate paths of the server ports 80 (HTTP) and 443 (HTTPS).
\end{itemize}
For further detail check the online repository%
\footnote{\url{https://github.com/hacid-project/climate-services-kg-docker}}.

\section{Functional Evaluation}\label{sec:eval}

As described in Section~\ref{sec:meth-xd}, eXtreme Design has been used for the development and evaluation of the ontology and the knowledge graph, considering the methodology's modularity and test-driven features. The defined competency questions (see Section~\ref{sec:reqs}) have been used to evaluate both the TBox (the ontology) and the ABox (the actual data in the knowledge graph) by means of the Competency Question Verification test. The ability to design SPARQL queries that address competency questions demonstrates the completeness of the ontology. Moreover, by running those queries on the knowledge graph and obtaining meaningful results, the salience and coherence of the knowledge graph are shown. 

\begin{figure}[htbp]
    \centering
 %   \begin{minipage}{0.9\textwidth}
% PREFIX rdfs: <http://www.w3.org/2000/01/rdf-schema#>
% PREFIX top: <https://w3id.org/hacid/onto/top-level/> 
% PREFIX ccso: <https://w3id.org/hacid/onto/ccso/>
% PREFIX data: <https://w3id.org/hacid/onto/data/>
% PREFIX rcp: <https://w3id.org/hacid/data/cs/greenhousegasconcentrationpathway/>
% PREFIX mip: <https://w3id.org/hacid/data/cs/variable/mip/>
% PREFIX dim: <https://w3id.org/hacid/data/cs/dimension/>

    \lstset{language=SPARQL, frame=single, basicstyle=\ttfamily\scriptsize}
    \caption{Example SPARQL query to filter simulations (prefix declarations omitted for brevity)}
    \label{sparql:filter_simulations}
    \begin{lstlisting}
# Filter available simulations by chosen emission scenario (in this case
# RCP4.5), available variables (in this case tas), and required spatial
# resolution (in this case less than 0.2 degrees).

SELECT ?model ?simulation ?output ?geodeticResolution
WHERE {
    ?simulation a ccso:Simulation ;
        ccso:refersToScenario rcp:RCP4.5 ;
        ccso:hasOutput ?output .
    ?output data:holdsSpecializationOfVariable* mip:tas ;
    data:dependsOnVariable ?geodeticVariable .
    ?geodeticVariable
        data:holdsSpecializationOfVariable* dim:geodetic ;
        data:hasDiscretization ?geodeticDiscretization .
    ?geodeticDiscretization a data:RollingRegularGrid ;
    	data:hasResolutionValue ?geodeticResolution .
	FILTER(?geodeticResolution < 0.2)
}
ORDER BY ?model ?simulation ?output
    \end{lstlisting}
 %   \end{minipage}
\end{figure}
\begin{figure}[htbp]

%    \vspace{1em}

%    \begin{minipage}{0.9\textwidth}
% PREFIX xsd: <http://www.w3.org/2001/XMLSchema#>
% PREFIX rdfs: <http://www.w3.org/2000/01/rdf-schema#>
% PREFIX top: <https://w3id.org/hacid/onto/top-level/> 
% PREFIX ccso: <https://w3id.org/hacid/onto/ccso/>
% PREFIX data: <https://w3id.org/hacid/onto/data/>
% PREFIX mip: <https://w3id.org/hacid/data/cs/variable/mip/>
% PREFIX dimension: <https://w3id.org/hacid/data/cs/dimension/>
% PREFIX sim: <https://w3id.org/hacid/data/cs/simulation/>
    \lstset{language=SPARQL, frame=single, basicstyle=\ttfamily\scriptsize}
    \caption{Example SPARQL query to filter output datasets (prefix declarations omitted for brevity)}
    \label{sparql:filter_output_datasets}
    \begin{lstlisting}
# Filter datasets from a simulation (in this case cmip5.HadCM3.rcp45.r10i1p1),
# based on available variables (in this case tas), and required exact temporal
# resolution (in this case one month).

SELECT ?dataset
WHERE {
    sim:cmip5.HadCM3.rcp45.r10i1p1
        ccso:hasOutput/top:hasPart* ?dataset.
    ?dataset data:holdsSpecializationOfVariable* mip:tas ;
    	data:dependsOnVariable ?temporalVariable .
    ?temporalVariable
        data:holdsSpecializationOfVariable* dimension:time ;
        data:hasDiscretization ?temporalDiscretization .
    ?temporalDiscretization a data:RollingRegularGrid ;
    	data:hasResolutionValue "P1M"^^xsd:duration .
}
ORDER BY ?dataset
    \end{lstlisting}
%    \end{minipage}
\end{figure}

SPARQL queries have been developed for all CQs.
Figures \ref{sparql:filter_simulations} and \ref{sparql:filter_output_datasets} contain SPARQL queries realizing the two example CQs shown in Section~\ref{sec:reqs}.

\section{Conclusions}\label{sec:conclusions}

In this paper, we have introduced the Climate Change Knowledge Graph, a significant step forward in the integration and accessibility of climate model data.
This resource aims %innovative framework
to address the limitations of current data retrieval systems %employed in the domain 
by providing a coherent, interoperable resource that facilitates complex queries across a wide spectrum of climate data.
%ur knowledge graph systematically aggregates diverse datasets and metadata, transforming how experts can access, analyze, and interpret climate simulations and their underlying models.
By engaging closely with domain experts throughout the development process, we ensured that the ontology and data representation are aligned with the practical needs and expectations of the climate science community, especially with respect to the process involved in the provisioning of climate services.
%This iterative approach has facilitated the creation of a tool that not only has the potential to streamline research processes but also enhances the ability of scientists, policymakers, and other stakeholders to make informed decisions about climate change mitigation and adaptation strategies.
The knowledge graph has an open access license and it is accessible online in multiple ways, to maximize interoperability.

%Our work demonstrates that knowledge graphs are powerful tools for managing complexity in data-rich fields like climate science. By illustrating the potential applications and advantages of this new system through various use cases, the Climate Change Knowledge Graph establishes a foundation for future advancements in climate informatics.

The knowledge graph is a living artefact and some new work is already under way. In particular climate experts expressed interest in having a way to represent the details of the process of climate service provisioning, so that instances of such process can be compared, shared, used to build collaborative solutions.
A description of the process in being built as a set of controlled vocabularies of roles, plans, tasks, and methods that address different aspects of the process.
%Furthermore, controlled vocabularies for methods used in the process are being developed (e.g., methods that can be used for bias correction).
% currently being extended with a controlled vocabulary that categorizes the methods that are used for executing several tasks in the Climate Service Workflow ontology. It includes three categories of methods, which are \textit{Spatial Downscaling}, \textit{Bias Correction}, and \textit{Statistical Analysis}. The controlled vocabulary serves two main functions: 1) it is a means of extracting granular information from the case contributor, and 2) it can be used for the evaluation of the contribution. 

%Looking forward, several pathways for future work are evident. First, the expansion of the knowledge graph to include additional datasets, including those outside the domain of climate models, could further enrich its utility. Second, the implementation of advanced analytics and machine learning techniques could enable users to derive even deeper insights from the available data. Finally, fostering collaborations with broader communities in environmental science and policy could ensure the continued relevance and impact of this tool.

%The Climate Change Knowledge Graph represents a significant advancement in facilitating the comprehensive exploration of climate data. By simplifying access to complex datasets and supporting in-depth analyses, it positions researchers and decision-makers to better understand and respond to the multifaceted challenges posed by climate change.

% Future work

\begin{credits}
\subsubsection{\ackname} 
This work was supported by the European Union through the Horizon Europe research and innovation programme within the context of the project HACID (Hybrid Human Artificial Collective Intelligence in Open-Ended Domains, grant agreement No 101070588) and the NextGenerationEU programme within the context of the project FOSSR (Fostering Open Science in Social Science Research) under NRRP Grant agreement n. MUR IR0000008. 
\end{credits}

%
% ---- Bibliography ----
%
% BibTeX users should specify bibliography style 'splncs04'.
% References will then be sorted and formatted in the correct style.
%
\bibliographystyle{splncs04}
\bibliography{bibliography}
\end{document}